\documentclass{ws-mpla}

\begin{document}

\catchline{}{}{}{}{}

\title{Precise Determinations of the Charm and Bottom Quark Masses}

\author{\footnotesize Sebastian Bodenstein}
\address{Department of Physics, University of Cape Town, Address\\
Cape Town, Rondebosch 7700, South Africa\\
sebastianbod@gmail.com}
\maketitle

\begin{abstract}
A finite-energy sum-rule is presented that allows for the use of combinations of both positive- and inverse-moment integration kernels. The freedom afforded from being able to employ this large class of integration kernels in our sum-rule is then exploited to obtain the values of the charm and bottom masses with minimum total uncertainty. We obtain as our final results $\bar{m}_c(3\,\text{GeV})=986(13)\,\text{MeV}$ and $\bar{m}_b(10\,\text{GeV})=3617(25)\,\text{MeV}$, which are amongst the most precise values of these parameters obtained by any method.

\end{abstract}

\section{Introduction}
There exist two competitive approaches for precision charm- and bottom-quark mass determinations. The first approach involves solving QCD on the lattice (LQCD). This is achieved by simulating the pseudoscalar correlator on the lattice and comparing it with the four-loop perturbative prediction.\cite{LQCD} This approach yields for the charm and bottom quarks (in the $\overline{\text{MS}}$-scheme)
\begin{align}
\bar{m}_c(3\,\text{GeV})&=986(6)\,\text{MeV},\\
\bar{m}_b(10\,\text{GeV})&=3617(25)\,\text{MeV}.
\end{align}
The second approach is the method of QCD sum-rules, pioneered by the ITEP group.\cite{shifman} This approach takes as theoretical input the Operator Product Expansion and experimental input the measured $e^+e^-\to Q\bar{Q}$ cross-section, where $Q=\{c,b\}$ are either the charm or bottom quarks. An example of a recent determination\cite{chetyrkin2009} of the charm and bottom masses along these lines obtained
\begin{align}
\bar{m}_c(3\,\text{GeV})&=986(13)\,\text{MeV},\\
\bar{m}_b(10\,\text{GeV})&=3610(16)\,\text{MeV}.
\end{align}
These masses are in remarkable agreement with the LQCD results. The philosophy for obtaining the above sum-rule results above was to to consider the Hilbert-moment sum-rule
\begin{equation}\label{EQ:chet}
\int \frac{ds}{s^{n+1}}R_Q(s)=\frac{12\pi^2}{n!}\left(\frac{d}{ds}\right)^n\Pi_Q(s)\bigl|_{s=0},
\end{equation}
where $\Pi_Q(s)$ is the vector correlator of two $Q$-flavoured quark currents. The right-hand side can be calculated using the OPE, whilst the left-hand side is evaluated using experimental data and pQCD. The quark mass, which enters $\Pi_Q(s)$, is determined by demanding that Eq. \eqref{EQ:chet} is satisfied. The only freedom one has is to choose $n$, which controls the weight function over which the data is to be integrated. The optimal value of $n$ is chosen on the basis of producing the lowest total uncertainty in the mass.\\  However, the analyticity of the correlator $\Pi_Q(s)$ allows one to write down more general sum-rules, for which Eq. \eqref{EQ:chet} is only a special case. Such sum-rules allow for a far greater choice of weight functions over which to integrate the data. Such sum-rules were considered in our work Ref. \refcite{bodCharm,bodBottom}. The philosophy for choosing the weight function is then two-part: First, enumerate the uncertainties affecting our mass determination, and second, choose the weight function that minimizes the total uncertainty. Thus the determination of the quark masses has been turned into a non-trivial optimization problem. Such a procedure was indeed found to reduce the total uncertainties, finding as final results
\begin{align}
\bar{m}_c(3\,\text{GeV})&=987(9)\,\text{MeV}\\
\bar{m}_b(10\,\text{GeV})&=3623(9)\,\text{MeV}.
\end{align}
where we used identical error metrics as used by Ref. \refcite{chetyrkin2009}, though using slightly lower uncertainties for some input parameters. These are the lowest uncertainties for any sum rule approach for the heavy quark masses, the lowest uncertainty for any method for the bottom quark mass. \\ \\ The purpose of this work will be to give an introduction to this sum rule approach.

\section{The Method}
We first consider the vector current correlator
\begin{eqnarray}
\Pi_{\mu\nu} (q^2) &=& i \int d^4x\,  e^{iqx} \langle 0| T[V_\mu(x) V_\nu(0)]|0\rangle \nonumber\\ 
&=& (q_\mu q_\nu - q^2 g_{\mu\nu}) \Pi_Q(q^2),
\end{eqnarray}
where $V_\mu(x) = \bar{Q}(x) \gamma_\mu Q(x)$. As we are considering the charm and bottom quarks, we have $Q=\{c,b\}$. Unitarity of the correlator implies the Optical Theorem,  
\begin{equation}
R_Q(s)=12\pi\text{Im}\Pi_Q(s),
\end{equation} with $R_Q(s)$ the standard $R$-ratio for $Q$-quark production. As is well-known, causality implies the analyticity of the correlator $\Pi_Q(s)$ in the entire complex plane, except for a branch cut along the positive real axis. The Residue Theorem in the complex $s$-plane thus implies
\begin{equation}\label{EQ:sumrule}
\int_{0}^{s_0}p(s) R_Q(s)ds=6\pi i\oint_{C(|s_0|)} p(s)\Pi_Q(s)ds+12\pi^2\text{Res}[\Pi_Q(s)p(s),s=0],
\end{equation}
where $p(s)$ is an arbitrary Laurent polynomial. The left-hand side of Eq. \eqref{EQ:sumrule} will be evaluated using experimental data, whilst we calculate the right-hand side by assuming that the exact correlator can be replaced by the OPE prediction, $\Pi_Q(s)\to \Pi_Q(s)\bigl|_{\text{OPE}}$. Eq. \eqref{EQ:sumrule} is our key expression, which generalizes some other sum-rules found in the literature. For example, if $p(s)=s^{-n}$ for $n=2,3,4,\ldots$ and $s_0\to \infty$, the contour integral in Eq. \eqref{EQ:sumrule} vanishes and one obtains the Hilbert moment sum rule Eq. \eqref{EQ:chet}. If instead one takes $p(s)$ to be some polynomial (ie. only containing positive powers of $s$), then the residue is zero in Eq. \eqref{EQ:sumrule} and one obtains the finite energy sum rule used in Ref. \refcite{schilcher}.\\ $\Pi_Q(s)\bigl|_{\text{OPE}}$ can be parameterized as 
\begin{equation}
\Pi_Q(s)\bigl|_{\text{OPE}}=\Pi_Q(s)\bigl|_{\text{pQCD}}+\Pi_Q(s)\bigl|_{\text{NP}}+\Pi_Q(s)\bigl|_{\text{qed}},
\end{equation}
where $\Pi_Q(s)\bigl|_{\text{pQCD}}$ is the dominant pQCD contribution, $\Pi_Q(s)\bigl|_{\text{NP}}$ are non-perturbative contributions and $\Pi_Q(s)\bigl|_{\text{qed}}$ are QED contributions. \\ Crucial to achieving high-precision using this approach is the availability of $\mathcal{O}(\alpha_{s}^{3})$ results for $\Pi_Q(s)\bigl|_{\text{pQCD}}$ in both the high- and low-energy limits. The Laurent-expansion of $\Pi_Q(s)\bigl|_{\text{pQCD}}$ about $s=-\infty$ has the form
\begin{equation}\label{HEE}
\Pi_Q(s)\bigl|_{\text{pQCD}}=e_{Q}^{2}\sum_{n=0}\Bigl(\frac{\alpha_s(\mu^2)}{\pi}\Bigr)^n \Pi^{(n)}(s),
\end{equation}
where $e_{Q}$ is the $Q$-quark electric charge ($e_c=2/3$, $e_b=-1/3)$. We also have 
\begin{equation}
\Pi^{(n)}(s)=\sum_{i=0}\Bigl(\frac{\bar{m}_{Q}^{2}}{s}\Bigr)\Pi_{i}^{(n)}.
\end{equation}
Here $\bar{m}_Q\equiv\bar{m}_Q(\mu)$ is the $\overline{\text{MS}}$-scheme quark mass at the renormalization scale $\mu$. We take the $\mathcal{O}(\alpha_{s}^{2}(\bar{m}_{b}^{2})^6)$ result from Ref. \refcite{QCD1}. At $\mathcal{O}(\alpha^{3}_{s})$ we have $\Pi_{0}^{(3)}$ and $\Pi_{1}^{(3)}$ from Ref. \refcite{QCD2}, and the logarithmic terms for $\Pi_{2}^{(3)}$ from Ref. \refcite{QCD3}. The constant term in $\Pi_{2}^{(3)}$ is not known exactly, but has been been estimated using Pade approximants.\cite{QCD4} At $\mathcal{O}(\alpha^{4}_{s})$ we have the exact logarithmic terms for $\Pi_{0}^{(4)}$ and $\Pi_{1}^{(4)}$,\cite{QCD5,QCD6} whilst the constant terms are not yet known. Given that these constant terms will contribute when we use kernels containing terms $s^{-1}$ and $s^0$ respectively, we will for the sake of consistency not include any $\mathcal{O}(\alpha^{4}_{s})$ terms. \\
We can Taylor expand $\Pi_Q(s)$ about $s=0$, which is customarily cast in the form
\begin{equation}
\Pi_Q(s)\bigl|_{\text{pQCD}}=\frac{3 e^{2}_{Q}}{16\pi^2}\sum_{n\geq 0}\bar{C}_n z^n,
\end{equation}    
where $z\equiv s/(4\bar{m}_{Q}^{2})$. The coefficients $\bar{C}_n$ can be expanded in powers of $\alpha_s(\mu)$
\begin{align}
\bar{C}_n=&\bar{C}_{n}^{(0)}+\frac{\alpha_s(\mu)}{\pi}(\bar{C}_{n}^{(10)}+\bar{C}_{n}^{(11)}l_m)\nonumber\\
&+\Bigl(\frac{\alpha_s(\mu)}{\pi}\Bigr)^2(\bar{C}_{n}^{(20)}+\bar{C}_{n}^{(21)}l_m+\bar{C}_{n}^{(22)}l^{2}_{m})\\
&+\Bigl(\frac{\alpha_s(\mu)}{\pi}\Bigr)^3(\bar{C}_{n}^{(30)}+\bar{C}_{n}^{(31)}l_m+\bar{C}_{n}^{(32)}l^{2}_{m}\\
&+\bar{C}_{n}^{(33)}l^{3}_{m})+\ldots
\end{align}
where $l_m\equiv \ln(\bar{m}^{2}_Q/\mu^2)$. Up to $\mathcal{O}(\alpha_{s}^{2})$, the coefficients up to $n=30$ of $\bar{C}_n$ are known.\cite{QCD8,QCD9} At $\mathcal{O}(\alpha_{s}^{3})$, we have $\bar{C}_0$ and $\bar{C}_1$ from Ref. \refcite{QCD8,QCD10}, $\bar{C}_2$ from Ref. \refcite{QCD9}, and $\bar{C}_3$ from Ref. \refcite{QCD11}. The kernel $p(s)$ will be chosen so that no coefficients $\bar{C}_4$ and above contribute to our sum rule \eqref{EQ:sumrule}. \\ The leading-order contribution to $\Pi_Q(s)\bigl|_{\text{NP}}$ comes from the gluon condensate. This contribution has been determined in Ref. \refcite{gluon}. The contribution of the gluon condensate is negligible for the bottom quark mass, and contributes roughly $1\,\text{MeV}$ to the charm quark mass. 
\\ We will use the Particle Data Group value for the strong coupling, $\alpha_s(M_{Z})=0.1184(7)$.\cite{coupling}

\section{The Charm Quark Mass}
\begin{table}
\tbl{Results for $\bar{m}_c(3\,\mbox{GeV})$ using three different kernels. The total uncertainty ($\Delta$Total) is obtained by adding all of the uncertainties in quadrature. }
{\begin{tabular}{@{}ccccccccc@{}} \toprule
& &  & \multicolumn{4}{c}{Uncertainties (MeV)} &  \\
\cline{3-8}
\noalign{\smallskip}
$p(s)$ 	& $\bar{m}_c(3\,\mbox{GeV})$ & $\Delta\text{Exp.}$ & 	$\Delta \alpha_s$  &	 $\Delta \mu$		& NP	&	 	   $\Delta s_0$ &  $\Delta$Total   \\
\hline
\noalign{\smallskip}
$s^{-2}$ 			      &  	995  & 9 	& 3			& 1		& 1  &  14	& 17 	  \\
$1-s_0/s$ 					&  	995  & 9 	& 4			& 3		& 3  &  6		& 12  \\
$1-(s_0/s)^2$ 			&  	987  & 7 & 4		  & 2		& 1	 &  4	  & 9\\\ \botrule 
\end{tabular}\label{TAB:charm} }
\end{table}

\begin{figure}
\centering
\includegraphics[scale=0.7]{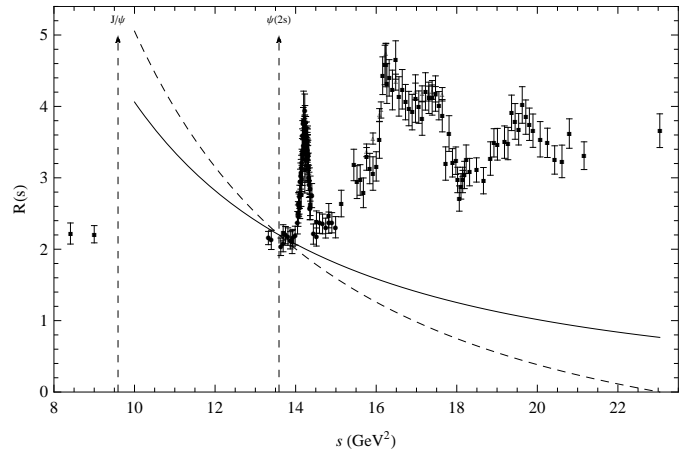}
\caption{The experimental data for $R_{udsc}(s)$ along with the kernel $p(s)=s^{-2}$ (solid line) and $p(s)=1-(s_0/s)^2$ (dashed line).}\label{fig:charmdata}
\end{figure}

\begin{figure}
\centering
\includegraphics[scale=0.7]{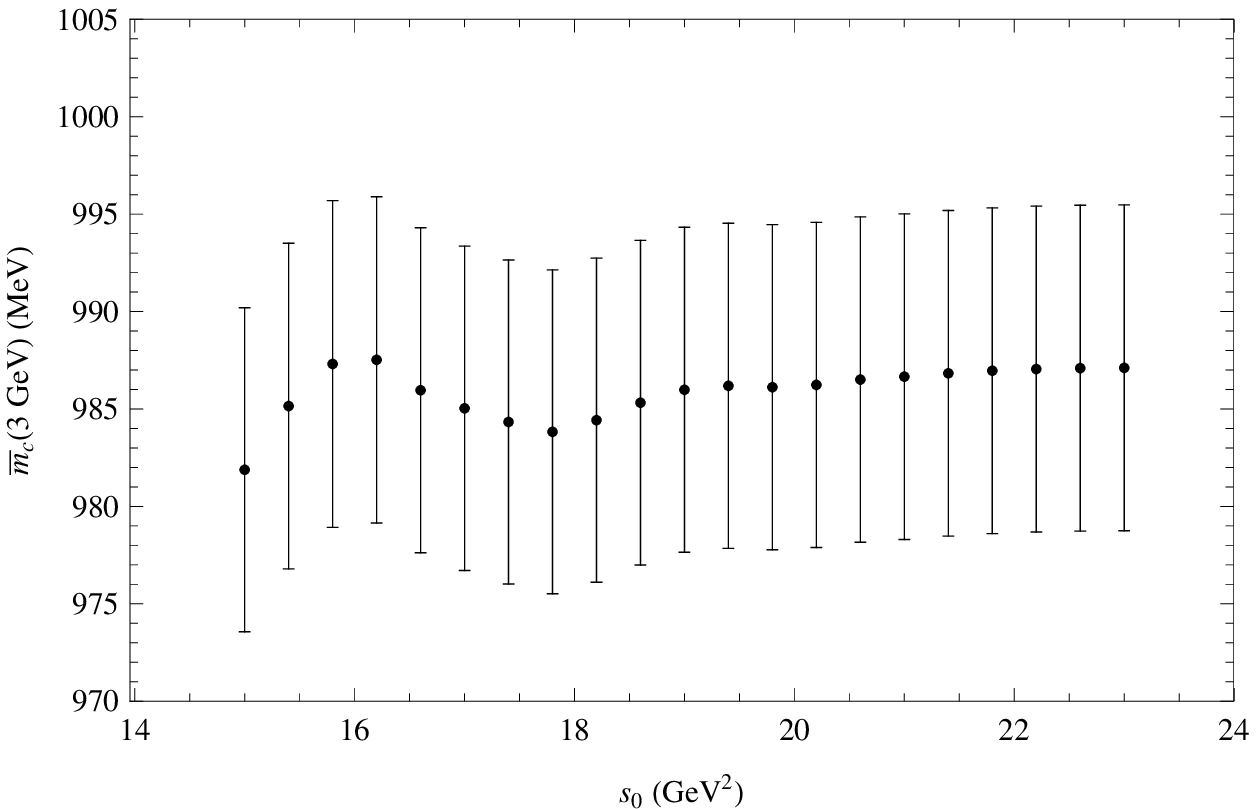}
\caption{The result for $\bar{m}_c(3\,\text{GeV})$ versus $s_0$ obtained using the kernel $p(s)=1-(s_0/s)^2$.}\label{fig:stab1}
\end{figure}

\begin{figure}
\centering
\includegraphics[scale=0.7]{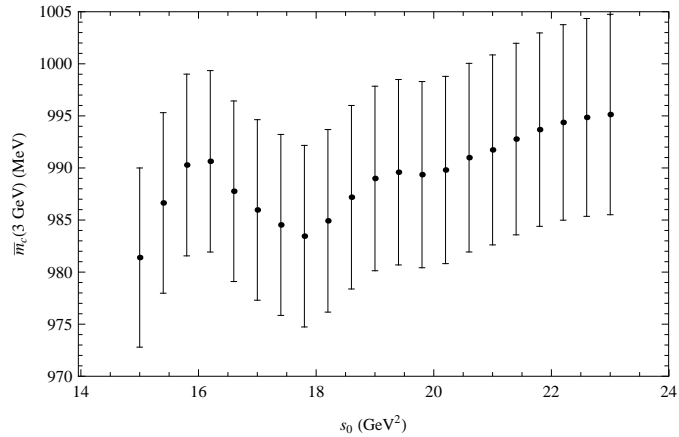}
\caption{The result for $\bar{m}_c(3\,\text{GeV})$ versus $s_0$ obtained using the kernel $p(s)=s^{-2}$.}\label{fig:stab2}
\end{figure}

Let us now consider the bottom quark mass. For the data input in Eq. \eqref{EQ:sumrule}, we used the zero-width approximation for the $J/\psi$ and $\psi(2s)$ resonances. Their masses and widths are taken from the Particle Data Group,\cite{PDG} whilst the effective electromagnetic couplings are taken from Ref. \refcite{kuhn2007}. For the resonance threshold region, which we take to be the range $(M_{\psi(2s)},4.8\,\text{GeV})$, we use a combination of BES\cite{BES1,BES2,BES3} and CLEO\cite{CLEO} measurements of $R_{udsc}$. In order to obtain $R_{c}$ to use in our sum-rule Eq. \eqref{EQ:sumrule}, we subtracted the light contribution, $R_{uds}(s)$, from $R_{udsc}(s)$. We use the pQCD prediction of $R_{uds}(s)$.\\ \\ The kernel $p(s)$ is chosen on the basis that it minimizes the total uncertainty in the charm mass. The primary uncertainties come from the experimental data ($\Delta \text{EXP}$), the strong coupling ($\Delta\alpha_s$), variation of the renormalization scale by $50\%$ about $\mu=3\,\text{GeV}$ ($\Delta\mu$), and the gluon condensate ($\Delta \text{NP}$). Finally, we also want our method to insensitive to variations in our integration end-point $s_0$. We check this by varying $s_0$ in the range $15-23\,\text{GeV}^2$ ($\Delta s_0$). This gives a (conservative) estimate of the uncertainty due to us not being in the duality regime yet at $\sqrt{s_0}=4.8\,\text{GeV}$. \\  In the work Ref. \refcite{chetyrkin2009}, kernels of the form $p(s)=s^{-n}$ for $n>1$ were considered, with $n=2$ found to be optimal. Although higher powers of $n$ better suppress the low-precision continuum threshold data in favour of the higher-precision $J/\psi$ and $\psi(2s)$ resonance contributions, this comes at the expense of the convergence of the OPE. The gluon condensate starts to contribute more, the mass becomes more sensitive on the value of the strong coupling, as well as on variations on $\mu$. It was found that the optimal kernels were of the form $p(s)=1-(s_0/s)^n$ for $n\geq 1$. The $n=2$ case is compared to the kernel $p(s)=s^{-2}$ in Fig. \ref{fig:charmdata}, where it can be appreciated how much better this kernel is of suppressing the poorly known continuum threshold data. The results are given in Table \ref{TAB:charm}, with a full uncertainty breakdown. Our best result, obtained using $p(s)=1-(s_0/s)^2$, is 
\begin{equation}
\bar{m}_c(3\,\text{GeV})= 987(9)\,\text{GeV}.
\end{equation}
It should be noted that even if we do not account for $\Delta s_0$ uncertainties, this result is more precise than that obtained using $p(s)=s^{-2}$. We plot the stability under variations of $s_0$ of the mass obtained using $p(s)=1-(s_0/s)^2$ and $p(s)=s^{-2}$ in Fig. \ref{fig:stab1} and Fig. \ref{fig:stab2} respectively.

\section{The Bottom Quark Mass}
\begin{figure}
\centering
\includegraphics[scale=0.7]{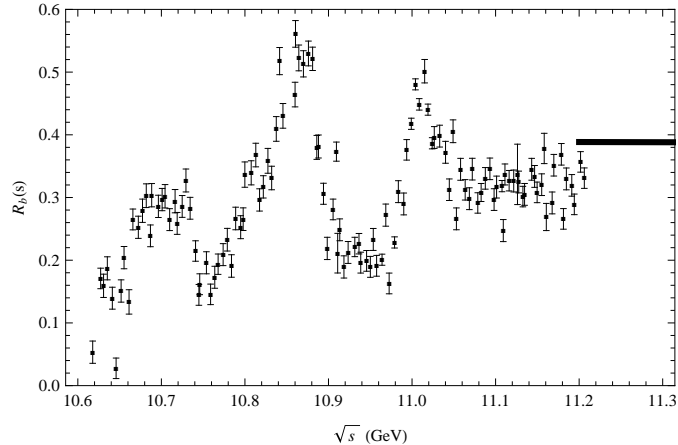}
\caption{The \emph{BABAR} data\protect\cite{BABAR} for $R_b$ corrected according to the prescription in Ref. \protect\refcite{chetyrkin2009}. The solid line is the pQCD prediction.}\label{fig:bottomdata}
\end{figure}

\begin{figure}
\centering
\includegraphics[scale=0.7]{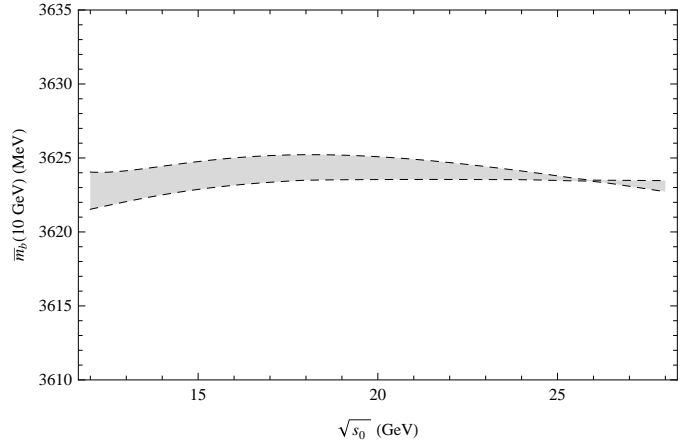}
\caption{The range of values obtained for $\bar{m}_b(10\,\text{GeV})$ using kernels of the form $\mathcal{P}_{3}^{(i,j,k)}(s,s_0)$ for varying $s_0$. }\label{fig:bottomstab}
\end{figure}

\begin{table}
\tbl{Results for $\bar{m}_b(10\,\text{GeV})$ using a variety of different kernels $p(s)$. The sources of uncertainties are from experiment ($\Delta\text{EXP}$), variation of the renormalization scale by $\pm 5\,\text{GeV}$ about $\mu=10\,\text{GeV}$ ($\Delta\mu$), and the strong coupling constant ($\Delta\alpha_s$). Finally, we include the uncertainties from calculating $\bar{m}_b(10\,\text{GeV})$ with and without \emph{Options A, B}, or \emph{C}. As in Refs. \protect\refcite{chetyrkin2009,chetyrkin2012}, these are not added to the total uncertainty and are listed only for comparison
purposes. }
{\begin{tabular}{@{}cccccccccc@{}} \toprule
& &  & \multicolumn{4}{c}{Uncertainties (MeV)} &  \multicolumn{3}{c}{Options A, B and C (MeV)}\\
\cline{4-7}
\cline{8-10}
\noalign{\smallskip}
$p(s)$ 	& $\bar{m}_b(10\,\mbox{GeV})$ & $\sqrt{s_0}\,(\text{GeV})$  & $\Delta\text{Exp.}$ & 	$\Delta \alpha_s$  &	 $\Delta \mu$			&	 $\Delta$Total	  &  $\Delta A$  &  $\Delta B$   & $\Delta C$    \\
\hline
\noalign{\smallskip}
$s^{-3}$ 			&  	3612  						&  $\infty$		  &		9	 		&		4	 	&		1	 &	10		 					& 		20	 				 		&	-17 & 16 \\
$s^{-4}$ 			&  	3626  						&  $\infty$		  &		7	 		&		5	 	&		6	 &	10		 					& 		12	 			 		&	-12 & 8 \\ 
$\mathcal{P}_{3}^{(-3,-1,0)}(s_0,s)$ 	&  	 3624 		& 16			&   6		 &		6	 								&		2	 &	9		 & 		1	 		&	-6 	& 0 \\
$\mathcal{P}_{3}^{(-3,-1,1)}(s_0,s)$ 	&  	 3624 		& 16			&   6		 &		6	 								&		2	 &	9		 & 		2	 		&	-7 	& 0 \\
$\mathcal{P}_{3}^{(-3,0,1)}(s_0,s)$ 	&  	 3624 		& 16			&   7		 &		6	 								&		2	 &	9		 & 		2	 		&	-7 	& 0 \\
$\mathcal{P}_{3}^{(-1,0,1)}(s_0,s)$ 	&  	 3625 		& 16			&   8		 &		5	 								&		4	 &	10		& 	4	 		&	-12 & 0 \\ 
$\mathcal{P}_{4}^{(-3,-1,0,1)}(s_0,s)$ 	&  	 3623 		& 20			&   6		 &		6	 							&		3	 &	9		& 	0	 		&	-4 	& 0 \\\ \botrule
\end{tabular}\label{TAB:bottom} }
\end{table}
For the data input in Eq. \eqref{EQ:sumrule}, we used the zero-width approximation for the $\Upsilon(1S), \Upsilon(2S), \Upsilon(3S)$ and $\Upsilon(4S)$ resonances. Their masses and widths are taken from the Particle Data Group,\cite{PDG} whilst the effective electromagnetic couplings are taken from Ref. \refcite{kuhn2007}. Finally, we employ the recent \emph{BABAR} measurement of $R_b$ in the continuum threshold region between $10.62\,\text{GeV}$ and $11.21\,\text{GeV}$.\cite{BABAR} As was pointed out in Ref. \refcite{chetyrkin2009}, this data cannot be directly be used in our sum-rule, as the initial state radiation and radiative tail of the $\Upsilon(4S)$ resonance must be removed, as well as the vacuum polarization taken into account. The procedure detailed in Ref. \refcite{chetyrkin2009} was followed to obtain the data entering our sum rule Eq. \eqref{EQ:sumrule}. The data This data is shown in Fig. \ref{fig:bottomdata}, with the pQCD prediction (obtained using the Fortran program {\ttfamily RHAD}\cite{RHAD}) included. As can be seen, the pQCD prediction is larger than the highest-energy data point. There are three possibilities to account for this fact:
\begin{itemize}
\item[(i)] \emph{Option A}: The \emph{BABAR} data are correct, but pQCD is only valid at some higher energy, for example $\sqrt{s}=13\,\text{GeV}$. Use a linear interpolation between the last data point $R_{b}^{\text{exp}}(11.21\,\text{GeV})=0.32$ and $R_{b}^{\text{pQCD}}(13\,\text{GeV})=0.377$, rather than the prediction from \texttt{Rhad}. 
\item[(ii)] \emph{Option B}: The pQCD prediction from \texttt{RHAD} is correct, but the \emph{BABAR} data are incorrect, and there is some unaccounted for systematic uncertainty. In this case, we multiply all the data by a factor of 1.21 to make the data consistent with pQCD. 
\item[(iii)] \emph{Option C}: The \emph{BABAR} data are correct, and pQCD starts at $\sqrt{s}=11.24\,\text{GeV}$. However, the pQCD prediction of \texttt{RHAD} is incorrect. The exact analytical form for $R_{b}^{\text{pQCD}}$ (rather than just expansions at low- and high-energies) is only known at tree-level and one-loop level. At $\mathcal{O}(\alpha_{s}^{2})$ already, the full anayltic result has to be reconstructed using Pad\`e approximants to patch together information about $\Pi_b(s)$ obtained at $\sqrt{s}=0, \sqrt{s}=4m_b$ and $\sqrt{s}=-\infty$. Both the Pad\`e method, and the reliance on pQCD results obtained at threshold ($\sqrt{s}=4m_b$) could introduce unaccounted systematic errors. As a measure of the methods dependence on the reconstructed correlator, we will replace the \texttt{RHAD} prediction of $R_{b}^{\text{pQCD}}$ with the high-energy expansion prediction Eq. \eqref{HEE}, which is closer to the experimental result for $R_{b}^{\text{exp}}(11.21\,\text{GeV})$. 
\end{itemize}
The first two options were considered in Ref. \refcite{chetyrkin2012}. \emph{Option C} is the least plausible of the options, but we include it for completeness. The kernel $p(s)$ in Eq. \eqref{EQ:sumrule} is chosen to minimize the dependence of the mass on \emph{Options A,B,C}, as well as reducing the dependence the uncertainties from the strong coupling, varying $\mu$ (we vary $\mu$ in the range $[5,15]\,\text{GeV}$) and the experimental data. The optimal kernels were found to be linear combinations of three to four powers of $s$ in the set $\mathcal{S}=\{s^{-3},s^{-2},s^{-1},1,s\}$, that are determined by demanding that the kernel obeys a global constraint that reduces the contribution of the problematic energy regions. As an example, the order 3 Laurent polynomial is of the form $p(s)=\mathcal{P}_{3}^{(i,j,k)}(s,s_0)=A(s^{i}+B s^j+C s^k)$, and is determined by   
\begin{equation}\label{EQ:constraint}
\int_{s^*}^{s_0} \mathcal{P}_{3}^{(i,j,k)}(s,s_0) s^{-n}\,ds=0,
\end{equation}
for $n\in \{0,1\}$. This determines $\mathcal{P}_{3}^{(i,j,k)}$ up to an irrelevant constant term. Including lower-powers of $s$ in the set $\mathcal{S}$ produces poor convergence of the pQCD expansion, whilst including higher-powers of $s$ brings in unknown $\mathcal{O}(\alpha_{s}^{3})$ terms in the high-energy expansion. The value of $s^*$ is chosen to coincide with the end of the \emph{BABAR} data, $\sqrt{s^*}=11.21\,\text{GeV}$. It is clear that this choice of kernel will reduce the dependence on \emph{Option A} and \emph{Option C}. However, this kernel has the virtue of also diverging rapidly outside of the range $[s^*,s_0]$, which leads to a suppression of the potentially problematic \emph{BABAR} continuum threshold data in favour of the mroe precisely known narrow upsilon resonances. This will reduce the dependence on \emph{Option B}.\\ \\ The results for some selected kernels, with a full uncertainty breakdown, are shown in Table \ref{TAB:bottom}. As can be seen, our new approach gives very similar uncertainties to the popular kernels $p(s)=s^{-3}$ and $p(s)=s^{-4}$ when only considering the usual uncertainty metrics ($\Delta \alpha_s, \Delta\mu, \Delta \text{Exp}$). However, our method is far less sensitive to \emph{Options A,B,C}. In particular, the work Ref. \refcite{chetyrkin2009} obtained their final result using the kernel $p(s)=s^{-3}$, which is clearly highly sensitive to \emph{Options A,B,C}.  \\ \\ There are ${{5}\choose{3}}=10$ possible kernels in the class $\mathcal{P}_{3}^{(i,j,k)}(s,s_0)$ and ${{5}\choose{4}}=5$ different kernels in the class $\mathcal{P}_{4}^{(i,j,k,r)}(s,s_0)$. Each of these kernels put very different emphasis on the low- and high-energy pQCD expansions, and leads to significantly different sensitivities to \emph{Options A,B,C}. The range of mass values obtained using all 10 kernels in the class $\mathcal{P}_{3}^{(i,j,k)}(s,s_0)$ is plotted for varying $s_0$ in Fig. \ref{fig:bottomstab}. As can be seen, all of the values of the mass lie in the range $3621\,\text{MeV}\leq \bar{m}_b(10\,\text{GeV}\leq 3625\,\text{MeV}$ for a range of $12\,\text{GeV}<\sqrt{s_0}<28\,\text{GeV}$. Using instead the 5 kernels in the class $\mathcal{P}_{4}^{(i,j,k,l)}(s,s_0)$ over a range of $18\,\text{GeV}<\sqrt{s_0}<40\,\text{GeV}$, we obtain a mass range of $3621\,\text{MeV}\leq \bar{m}_b(10\,\text{GeV}\leq 3625\,\text{MeV}$, exactly the same as using the class $\mathcal{P}_{4}^{(i,j,k,r)}(s,s_0)$. This consistency of our mass result with using very different kernels over a very large range of $s_0$ inspires great confidence in this method. We give as our final result
\begin{equation}
\bar{m}_b(10\,\text{GeV})= 3623(9)\,\text{MeV}
\end{equation}
The total uncertainty is the uncertainty from varying $\mu$, the strong coupling, and experiment all added in quadrature. If we include the uncertainties from \emph{Options A,B,C}, the uncertainty in our result increases only marginally to $\bar{m}_b(10\,\text{GeV})= 3623(10)\,\text{MeV}$. This is in contrast to using the kernel $p(s)=s^{-3}$, whose uncertainty in the mass would more than double when including the \emph{Options A,B,C} uncertainty.

\section{Conclusion}
We introduced a new sum-rule Eq. \eqref{EQ:sumrule}, which allows us to employ both positive and inverse moments for the weight function $p(s)$ over which the experimental data is integrated. The results 
\begin{align}
\bar{m}_c(3\,\text{GeV})&=987(9)\,\text{MeV}\\
\bar{m}_b(10\,\text{GeV})&=3623(9)\,\text{MeV},
\end{align}
were obtained using the kernels $p(s)=1-(s_0/s)^2$ for the charm case and $p(s)=\mathcal{P}_{4}^{(-3,-1,0,1)}(s_0,s)$ in the bottom case. When our uncertainty metrics are from experiment, the strong coupling, variations in $\mu$ and from the gluon condensate, then these new kernels produce only modest uncertainty reductions in comparison to kernels of the form $p(s)=s^{-n}$. However, these new kernels truly excel when other systematic uncertainties need to be controlled.

\section*{Acknowledgments}
I would like to thank H. Spiesberger, C.A. Dominguez and K. Schilcher for a fruitful collaboration. I would like to thank the Institute of Theoretical Physics, Nanyang
Technological University, for their great hospitality during the conference.\\ This work was supported in part by NRF (South Africa) and Alexander von Humboldt Foundation (Germany), as well as the Post-Graduate Funding Office at the University of Cape Town.

\end{document}